\documentclass{llncs}

\usepackage{paradise}
\usepackage{longtable,booktabs}
\usepackage{graphicx}
\usepackage[sectionbib]{natbib}
\setcitestyle{numbers,square,comma}

\newcommand{\hypertarget}[2]{#2}

\usepackage{caption}
\begin{document}
\bibliographystyle{plainnat}

\pagestyle{plain}
\mainmatter

\title{Symbolic Computation via Program Transformation\thanks{This work has been partially supported by the Czech Science Foundation grant 18-02177S and by Red Hat, Inc.}}

\author{Henrich Lauko \and Petr Ročkai \and Jiří Barnat}
\institute{ \{xlauko1,xrockai,barnat\}@fi.muni.cz}

\maketitle

\begin{abstract}
  Symbolic computation is an important approach in automated program analysis. Most state-of-the-art tools perform symbolic computation as interpreters and directly maintain symbolic data. In this
  paper, we show that it is feasible, and in fact practical, to use a compiler-based strategy instead. Using compiler tooling, we propose and implement a transformation which takes a standard program
  and outputs a program that performs semantically equivalent, but partially symbolic, computation. The transformed program maintains symbolic values internally and operates directly on them hence the
  program can be processed by a tool without support for symbolic manipulation.
  
  The main motivation for the transformation is in symbolic verification, but there are many other possible use-cases, including test generation and concolic testing. Moreover using the transformation
  simplifies tools, since the symbolic computation is handled by the program directly. We have implemented the transformation at the level of \llvm{} bitcode. The paper includes an experimental evaluation,
  based on an explicit-state software model checker as a verification backend.
\end{abstract}

\section{Introduction}\label{introduction}

It is common to use symbolic methods in program analysis and verification and related disciplines. Symbolic execution has found numerous use cases in test generation and concolic testing and is widely
deployed in practice. Likewise, many modern software verification tools are based on bounded model checking, which combines symbolic execution with SMT solvers to successfully attack hard problems in
their problem domain.

On one hand, multiple production-quality SMT solvers are readily available and even provide a common interface~\citep{barrett16:satisf.modulo}. While a certain degree of integration is required to
achieve optimal performance, solvers have attained nearly commodity status. This is in stark contrast to symbolic interpretation, which is usually implemented ad-hoc and is not re-usable across tools
at all. The only exception may be KLEE~\citep{cadar08:klee}, a symbolic interpreter for \llvm{} bitcode, which is used as a backend by a few analysis tools. Undoubtedly, the fact that it is based on the
(ubiquitous) \llvm{} intermediate language has helped it foster wider adoption.

Arguably, interpreters (virtual machines) for controlled program execution, as required by dynamical analysis tools, are already complex enough, without involving symbolic computation. To faithfully
interpret real-world programs, many features are required, including an efficient memory representation, support for threads, exceptions and a mechanism to deal with system calls. Complexity is,
however, undesirable in any system and even more so in verification tools.

\begin{figure}
\centering
\includegraphics{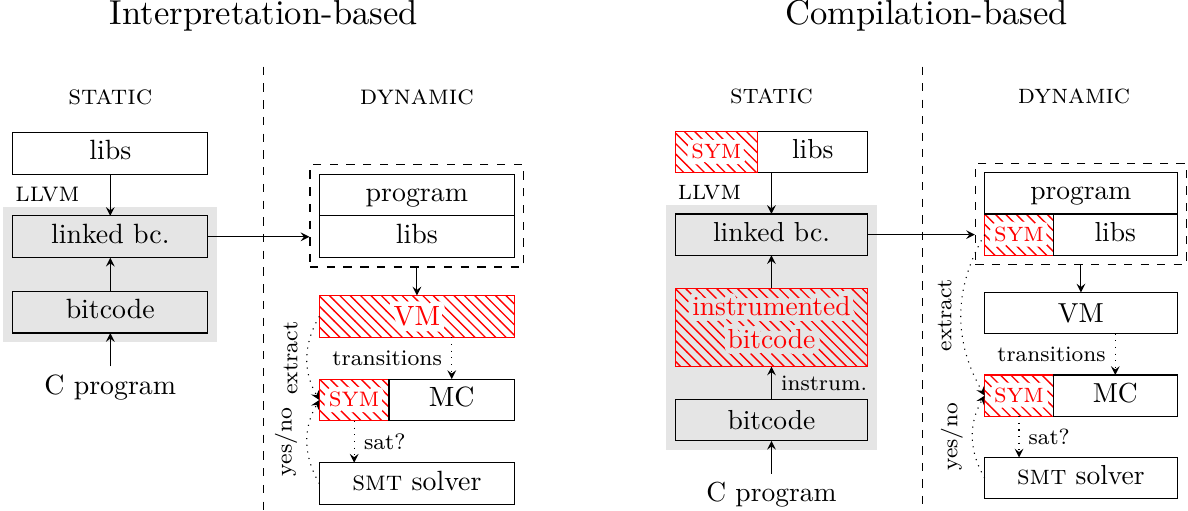}
\caption{Comparison of interpretation-based and compilation-based symbolic methods in the context of \llvm{} model checking. VM stands for `virtual machine', while MC stands for `model checker'. The
hatched boxes represent components that work with symbolic data. In the compilation-based method, symbolic operations are instrumented into the bitcode, and their implementation is provided in the
form of a library. The virtual machine does not need to know about symbolic values at all. The model checker, however, extracts symbolic data and a path condition from the executed
program.}\label{fig:compilation}
\end{figure}

For these reasons, we propose to lift symbolic computation into a separate, self-contained module with minimal interfaces to the rest of the verification or analysis system (see
Figure~\ref{fig:compilation}). The best way to achieve this is to make it \emph{compilation-based}, that is, provide a transformation that turns ordinary (explicit) programs into symbolic programs
automatically. The transformed program only uses explicit operations, but it uses them to manipulate symbolic expressions and as a result can be executed using off-the-shelf components.

The expected result is that the proposed transformation can be combined with an existing solver and a standard explicit interpreter of \llvm{} bitcode. Depending on how one combines those ingredients,
they will obtain different analysis tools. As an example, in Section~\ref{sec:symmc}, we use the transformation, an existing explicit-state model checker \divine{} and an SMT solver
STP~\citep{ganesh07:decision.proced.bit} to build a simple control-explicit, data-symbolic (CEDS)~\citep{bauch16:control.explic} model checker. Building a tool which implements symbolic execution
would be even simpler.\footnote{In fact, a control-explicit, data-symbolic model checker already contains a subroutine (in our case about 200 lines) which effectively implements a symbolic executor.}

\subsection{Goals}\label{sec:goals}

Our primary goal is to design a self-contained program transformation that can be used in conjunction with other components to piece together symbolic analysis and verification tools. We would like
the transformation to exhibit the following properties:

\begin{enumerate}
\def\labelenumi{\arabic{enumi}.}
\tightlist
\item
  allow mixing of explicit and symbolic computation in a single program,
\item
  expose a small interface to the rest of the system, and finally
\item
  impose minimal run-time overhead.
\end{enumerate}

The first property is important because it often does not make sense to perform all computation within a program symbolically. For instance, a symbolic execution engine may wish to natively perform
library calls requested by the program. Therefore, it ought to be possible to request, from the outset, that a particular value in the program is symbolic or explicit.

It is unfortunately not possible to execute the symbolised program in a context that is completely unaware of symbolic computation. However, the requirements imposed on the execution environment can
be minimised and defined clearly (see Section~\ref{sec:interfaces}). Finally, exploring all possible executions given a single input sequence is already expensive and when used in the context of model
checking, we would like to incur as small a penalty as possible.

\subsection{Contribution}\label{contribution}

The idea that various tasks can be shifted between compile time and run time is as old as higher-level programming languages. In the context of verification, there is a large variety of approaches
that put different tasks at different points between these two extremes. Symbolic computation is traditionally found near the \emph{interpretation} end of the spectrum.

Our contribution is to challenge this conventional wisdom and show that this technique can be shifted much farther towards the \emph{compilation} end. Further, by treating symbolic computation as an
\emph{abstract domain}, we pave the way for other abstract domains to be approached in this manner. Finally, all relevant source code and benchmark data is freely available online.\footnote{\url{https://divine.fi.muni.cz/2018/sym}}

\section{Related Work}\label{related-work}

Program verification techniques based on symbolic execution~\citep{king76:symbol.execut}, symbolic program code analysis~\citep{nielson99:princip} and symbolic approach to model
checking~\citep{mcmillan93:symbol.model.checkin} have been the subject of extensive research.

As for symbolic execution, the approach most closely related to ours is represented by the KLEE symbolic execution engine~\citep{cadar08:klee} that performs symbolic execution on top of \llvm{}
IR~\citep{lattner04:llvm}. Besides standalone usage as a symbolic executor, KLEE has become also a back-end tool for other types of analyses and for verification. For example, the tool
Symbiotic~\citep{chalupa17:symbiot} combines code instrumentation and slicing with KLEE to detect errors in C programs.

Besides symbolic execution, other forms of abstract interpretation, like predicate abstraction, is often used in code analysis. The most successful approaches are based either on counterexample-guided
abstraction refinement (CEGAR)~\citep{clarke00:counter.guided} or lazy abstraction with interpolants~\citep{albarghouthi12:from.under}, which are implemented in tools such as
BLAST~\citep{beyer07:blast} and CPAchecker~\citep{beyer11:cpachec}. There are numerous research results in this direction, summarised in
e.g.~\citep{beyer15:interp.value.analys, sousa17:abstrac.interp.unfold, weissenbacher10:program.analys.interp}.

A verification algorithm that goes beyond static program code analysis and combines predicate abstraction with concrete execution and dynamic analysis has been also introduced~\citep{daniel15:panda}.
This approach can successfully verify programs that feature unbounded loops and recursion, unlike standard symbolic execution.

Using instrumentation (as opposed to interpretation) for symbolic verification was proposed a few times, but the only extant implementation that works with realistic programs is derived from the
CUTE~\citep{sen05:cute} family of concolic testing systems, i.e.~the tools CREST~\citep{burnim08:heuris.scalab.dynamic} and jCUTE~\citep{sen06:cute}. In particular, CREST uses the CIL
toolkit\footnote{CIL is short for C Intermediate Language~\citep{necula02:cil}, and is a simplified subset of the C language. The toolkit can automatically translate standard C into the intermediate
  (CIL) form. The CIL form can be optionally brought into the form of three-address code and this feature is used in CREST.} to insert additional calls into the program to perform the symbolic part of
concolic execution. The approach as described in~\citep{sen05:cute} is limited to symbolic computation, unlike the present paper, which works with arbitrary abstract domains.

A related, process was proposed by Khurshid et al.~\citep{khurshid03:general.symbol}: in this case, hand-annotated code was processed by Java PathFinder~\citep{havelund00:model.java}, an explicit
state model checker. Our approach, in contrast, is fully automatic and more general.

Finally, besides symbolic code analysis and symbolic execution, there are approaches that perform symbolic model checking as such. The key differentiating aspect of symbolic model checking is the
ability to decide equality of symbolically represented states. This is important in particular for verification of parallel and reactive programs where the state space contains diamonds or loops,
respectively. The tool Sym\divine{}~\citep{mrazek16:symdiv} is focused on bit-precise symbolic model checking of parallel C and C++ programs. It extends standard explicit state space exploration with SMT
machinery to handle non-deterministic data values. As such, Sym\divine{} is halfway between a symbolic executor and an explicit-state model checker. Unlike the solution presented in this paper, Sym\divine{}
does not separate the symbolic interpreter from the core of the model checker. In general, symbolic model checking is more often used with synchronous systems, for
example~\citep{cavada14:symbol.model.checker}.

\section{Abstraction as a Transformation}\label{abstraction-as-a-transformation}

While in the present work, our main goal is to transform a concrete program into one that performs symbolic computation, it is expedient to formulate the problem more generally. We will think in terms
of an \emph{abstraction}, in the sense of abstract interpretation, which has two main components: it affects how \emph{program states} are represented and it affects the \emph{computation of
transitions} between those states. There are two levels on which the abstraction operates:

\begin{enumerate}
\def\labelenumi{\arabic{enumi}.}
\tightlist
\item
  the static level, concerning syntactic constructs and the type system
\item
  dynamic, or semantic, which concerns actual execution and values
\end{enumerate}

In the rest of this section, we will define \emph{syntactic abstraction} (which covers the static aspects) as means of encoding abstract semantics into a concrete program. While it is convenient to
think of the transformed program in terms of abstract values and abstract operations, it is also important to keep in mind that at a lower level, each abstract value is concretely represented
(encoded). Likewise, abstract operations (instructions) are realised as sequences of concrete instructions which operate on the concrete representation of abstract values (see
Figure~\ref{fig:simpledom}, left). Those considerations are at the core of the second, dynamic, aspect of abstraction. Reflecting this structure, the program transformation therefore proceeds in two
steps:

\begin{enumerate}
\def\labelenumi{\arabic{enumi}.}
\tightlist
\item
  the input program is (syntactically) abstracted

  \begin{itemize}
  \tightlist
  \item
    concrete values are replaced with abstract values
  \item
    concrete instructions are replaced with abstract instructions
  \end{itemize}
\item
  abstract instructions are replaced by their concrete realisation
\end{enumerate}

The remainder of this section is organised as follows: in Section~\ref{sec:states}, we describe the expected concrete semantics of the input program. Section~\ref{sec:syntactic} then introduces
syntactic abstraction, Section~\ref{sec:values} deals with representation and typing of values in the abstracted program, Section~\ref{sec:instructions} goes on to describe the treatment of
instructions. Section~\ref{sec:domains} briefly discusses interactions of multiple domains within a program and finally Section~\ref{sec:constraints} gives an overview of relational abstract domains
that we use to perform a symbolic computation.

\subsection{States and Transitions}\label{sec:states}

We are interested in general programs, e.g.~those written in the C programming language. Abstraction is often described in terms of \emph{states} and \emph{transitions}. In case of C programs, a state
is described by the content of memory (including registers). Transitions describe how a state changes during computation performed by a given program. In this paper, we will use small-step semantics,
partly because the prototype implementation is based on \llvm{},\footnote{Programs in \llvm{} are in a partial SSA form, a special case of three-address code~\citep{aho07:compil}. Three-address code is
  essentially small-step semantics in an executable form.} and in part because it is a natural choice for describing parallel programs.

In this description, the transitions between program states are given by the effect of individual instructions on program state. Which instruction is executed and which part of the program state it
affects is governed by the source state. Our discussion of \emph{abstract transitions} will therefore focus on the effects of instructions: as an example, the \texttt{add} instruction obtains two
values of a specified bit width from some locations in the program state, computes their sum and stores the result to a third location.

\subsection{Syntactic Abstraction}\label{sec:syntactic}

The input program is given as a collection of functions, each consisting of a control flow graph where nodes are basic blocks -- each a sequence of non-branching instructions. Memory access is always
explicit: there are instructions for reading and writing memory, but memory is never directly copied, or directly used in computation. While this further restricts the semantics of the input program,
it is not at the expense of generality: programs can be easily put in this form, often using commodity tools.

\begin{figure}
\centering
\includegraphics{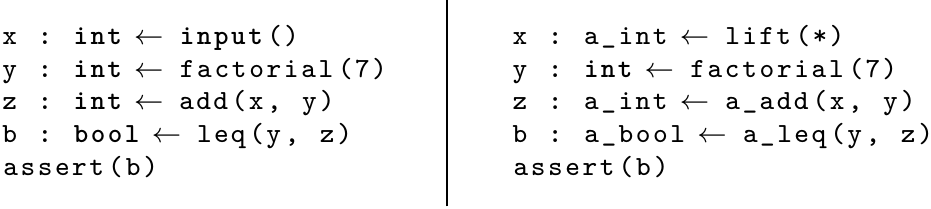}
\caption{An example of syntactic abstraction. In this example, \texttt{a\_int} and \texttt{a\_bool} represent abstract types (see also Section~\ref{sec:values}). We create the abstract value
\texttt{x} with a \texttt{lift(*)} operation to represent an arbitrary value of type \texttt{int} (see Section~\ref{sec:semantic}). Also, notice that the concrete computation of \texttt{factorial(7)}
remains intact.}\label{fig:syntactic}
\end{figure}

With these considerations in mind, the goal of what we will call \emph{syntactic abstraction} is to replace some of the concrete instructions with their abstract counterparts. The general idea is
illustrated in Figure~\ref{fig:syntactic}.

Apart from a few special cases, an abstract instruction takes abstract values as its inputs and produces an abstract value as its result. The specific meaning of those abstract instructions and
abstract values then defines the \emph{semantic abstraction}. The result of \emph{syntactic} abstraction being performed on the program is, therefore, that the modified program now performs abstract
computation. In other words, the transformed program directly operates on abstract states and the effect of the program on abstract states defines the abstract transition system.

We posit that syntactic abstraction, as explained in following sections, will automatically lead to a good semantic abstraction -- i.e.~one that fits the standard definition: a set of concrete states
can be mapped to an abstract state, an abstract state can be realised as a set of concrete states and those operations are compatible in the usual sense.

\subsection{Abstract Values and Static Types}\label{sec:values}

A distinguishing feature of the syntactic approach to abstraction is that it admits a static type system. In other words, the variables in the program can be assigned consistent types which respect
the boundary between abstract and concrete values. While a type system is a useful consistency check, its main importance lies in facilitating a description of how syntactic abstraction
operates.\footnote{Additionally, since the SSA portion of the \llvm{} IR is already statically typed, we can take advantage of this existing type system in the implementation. Nonetheless, the treatment
  in this section does not depend on \llvm{} and would be applicable to any dataflow-oriented program representation.}

We start by assuming existence of a set of \emph{concrete scalar types}, \(S\), and of concrete pointer types. We define a map \(Γ\) that builds up all relevant types from the set of scalar types. The
set of all types \(Γ(T)\) derived from a set of scalars \(T\) is defined inductively as follows:

\begin{enumerate}
\def\labelenumi{\arabic{enumi}.}
\tightlist
\item
  \(T ⊆ Γ(T)\), that is, each scalar type is included in \(Γ(T)\)
\item
  if \(t₁, ..., t_n ∈ Γ(T)\) then also the product type \((t₁, ..., t_n) ∈ Γ(T)\), \(n ∈ ℕ\)
\item
  if \(t₁, ..., t_n ∈ Γ(T)\) then also the disjoint union \(t₁ | t₂ | ... | t_n ∈ Γ(T)\), \(n ∈ ℕ\)
\item
  if \(t ∈ Γ(T)\) then \(t* ∈ Γ(T)\) (\(t*\) denotes a pointer type)
\end{enumerate}

In other words, the set \(Γ(S)\) describes finite (non-recursive) algebraic types over the set of concrete scalars and pointers.

A fundamental building block of the syntactic abstraction is a bijective map \(α_i\), defined for each abstract domain separately,\footnote{Since multiple abstract domains can co-exist in a single
  program, we use the lower index \(i\) to distinguish them.} from the set of \emph{concrete scalar types} \(S\) to the set of abstract scalar types \(A_i = α_i(S)\) (we let \(A\) be the union of all
the \(A_i\): \(A = A₁ ∪ A₂ ∪ ...\)). Each value which exists in the abstracted program then belongs to a type in \(Γ(S ∪ A)\) -- in other words, values are built up from concrete and abstract scalars.

In particular, this means that the abstraction works with \emph{mixed types} -- products and unions with both concrete and abstract fields. Likewise, it is possible to form pointers to both abstract
values and to mixed aggregates.

\subsection{Semantic Abstraction}\label{sec:semantic}

The maps \(α_i\) and \(α_i^{-1}\) let us move from concrete to abstract scalar \emph{types} (and back) and are strictly a syntactic construct. The \emph{semantic} (dynamic) counterpart of \(α_i\) are
\emph{lift}\(_i\) and \emph{lower}\(_i\): these are not maps, but rather abstract operations (instructions). Just as \(α_i\) and \(α_i^{-1}\) translate between concrete and abstract types,
\emph{lift}\(_i\) goes from concrete to abstract \emph{values} and \emph{lower}\(_i\) the other way around. While both the \(α_i\) and \emph{lift}\(_i\) and \emph{lower}\(_i\) are defined on scalar
types \(S\) and scalar values respectively, they can be all naturally extended to the set of all types \(Γ(S)\) (and their corresponding values).

\subsection{Representation}\label{representation}

Besides \(α_i\), there is another type map, which we will call \(ρ_i\) which maps each abstract scalar type in \(A_i\) to a concrete type in \(Γ(S)\). This is the \emph{representation map}, and
describes how abstract values are \emph{represented} at runtime. This is to emphasise that abstract values are, in the end, encoded using concrete values that belong to particular concrete types.
Moreover, in general for \(t ∈ Γ(S)\), \(ρ_i(α_i(t)) ≠ t\): the representation is unrelated to the original concrete type. An abstract floating point number may be, for instance, represented by a
concrete pointer to a concrete aggregate made of two 32-bit integers.

\begin{figure}
\centering
\includegraphics{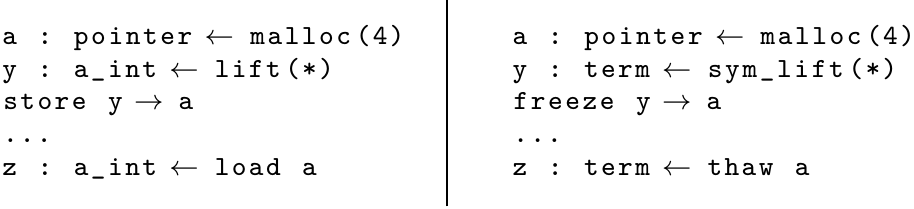}
\caption{Freezing and thawing of values transfers them between abstract representation and their concrete realisation. In this case, \(ρ\) sends \texttt{a\_int} to \texttt{term}, which realises the
term domain described in section Section~\ref{sec:symbolic}. The \emph{freeze} and \emph{thaw} operations allow \texttt{term} to be bigger than the original 4-byte integer
type.}\label{fig:freeze-thaw}
\end{figure}

While \emph{lift}\(_i\) and \emph{lower}\(_i\) are the value-level counterparts of the map \(α_i\), we need another pair of operations to accompany the representation map \(ρ_i\). We will call them
\emph{freeze}\(_i\) and \emph{thaw}\(_i\), and they map between \(t ∈ A_i\) and \(ρ_i(t) ∈ Γ(S)\). The idea is that memory manipulation (and manipulation of any concrete aggregates) is done entirely
in terms of the representation types (using frozen values), but abstraction operations on scalar values are defined in terms of the abstract type (i.e.~thawed values). The use of freezing and thawing
is illustrated in Figure~\ref{fig:freeze-thaw}.

One challenge in the implementation of \emph{freeze}\(_i\) and \emph{thaw}\(_i\) is that the memory layout of a program should not change\footnote{The exact layout of data (structures, arrays, dynamic
  memory) is normally the responsibility of the program itself, more so in the case of intermediate or low-level languages. For this reason, it is often the case that the program will make various
  assumptions about relationships among addresses within the same memory object. It is impractical, if not impossible, to automatically adapt the program to a different data layout, e.g.~in case the
  size of a scalar value would change due to abstraction.} as a side-effect of the transformation. This means that for many abstract domains, the \emph{freeze} operation must be able to store
additional data associated with a given address, and \emph{thaw} must be able to obtain this data efficiently. While this is an implementation issue, it is an important part of the interface between
the transformed program and the underlying execution or verification platform. However, since the program is transformed as a whole, there is no need to explicitly track this additional data at
runtime.\footnote{The only way a value can be copied from one memory address to another is via a \texttt{load} instruction followed by a \texttt{store}, both of which are instrumented and as such also
  transfer the supplementary data.}

An additional role of the \emph{freeze}/\emph{thaw} pair is to maintain dynamic type information at runtime. While it is easy to assign static types to instruction operands and results, this is not
true for memory locations: different parts of the program can load values of different static types from the same memory address. For this reason, the type system which governs memory use must be
dynamic and allow dispatch on the actual (runtime) type of the value stored at a given memory location.

\subsection{Abstract Instructions}\label{sec:instructions}

As indicated at the start of this section, it is advantageous to formulate the transformation in two phases, using intermediate abstract instructions. Abstract instructions take abstract values as
operands and give back abstract values as their results. It is, however, of crucial importance that each abstract instruction can be realised as a suitable sequence of concrete instructions. This is
what makes it possible to eventually obtain an abstract program that does not actually contain any abstract instructions and execute it using standard (concrete, explicit) methods.

In the first (abstraction) phase, concrete instructions are replaced with their abstract versions: instruction \texttt{inst} with a type \((t₁, ..., t_n) → t_r\) is replaced with \texttt{a\_inst} of
type \((α(t₁), ..., α(t_n)) → α(t_r)\). Additionally, \emph{lift}, \emph{lower}, \emph{freeze} and \emph{thaw} are inserted as required.\footnote{For instance, concrete operands to abstract operations
  are lifted, arguments to necessarily concrete functions (e.g.~real system calls) are lowered. Memory stores are replaced with \emph{freeze} and loads with \emph{thaw}.} The implementation is free to
decide which instructions to abstract and where to insert value lifting and lowering, as long as it obeys typing constraints. The specific approach we have taken is discussed in
Section~\ref{sec:domains} and the implementation aspects are described in Section~\ref{sec:impldomains}.

After the first phase is finished, the program may be further manipulated in its abstract form before continuing the second phase of the abstraction. This gives a practical, implementation-driven
reason for performing the abstraction transformation in two steps, in addition to the conceptual clarity it provides.

In the second step, all abstract operations, including \emph{lift} and \emph{lower}, are realised using concrete subroutines. The realisation (implementation) of \texttt{a\_inst} is of the type
\((ρ(α(t₁)), ..., ρ(α(t_n))) → ρ(α(t_r)))\), clearly obviating the need for thawing and re-freezing the value.

\subsection{Abstract Domains}\label{sec:domains}

Necessarily, in an abstracted program, the values it manipulates will come from at least two different domains: the concrete domain and the chosen abstract domain, in line with the first requirement
laid out in Section~\ref{sec:goals}. This is because it is usually impractical to abstract \emph{all} values that appear in the program. Additional abstract domains, therefore, do not pose any new
conceptual problems.

For the sake of simplicity, we only consider instructions where all operands come from the same domain (this holds for both the concrete and for abstract domains). Moreover, the only instructions
where the domain of the result does not agree with the domain of the operands are cross-domain conversion operations, which take care of transitioning values from one domain to another. The two most
important instances of those operations are \emph{lift} and \emph{lower}\footnote{The names \emph{lift} and \emph{lower} allude to the relationship of the abstract and the concrete domain. In
  applications with multiple abstract domains, it may be expedient to include additional instructions that convert directly from one abstract domain to another, although in theory it is always
  possible to go through the concrete domain.} introduced in Section~\ref{sec:values}.

\begin{figure}
\centering
\includegraphics{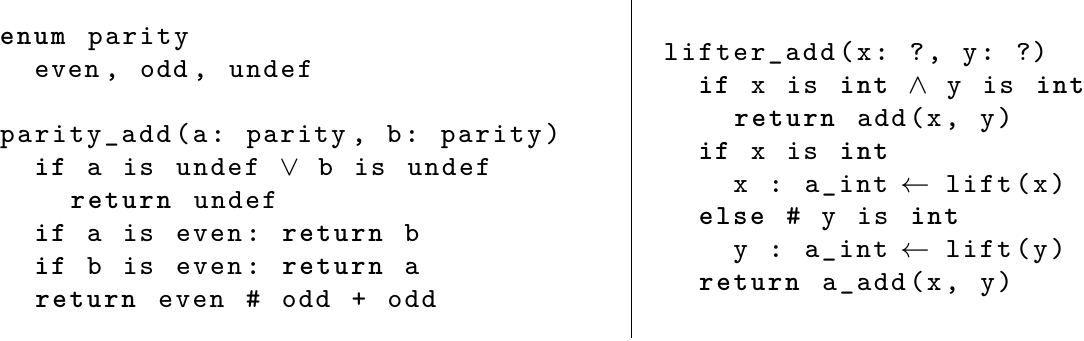}
\caption{Left: Domain implementation can be provided in a high-level language (e.g.~C++) and needs to provide a representation of abstract scalar values and operations on them. An abstract value (of
type \texttt{parity}) can be \texttt{even}, \texttt{odd} or in a superposition of those (\texttt{undef} -- unknown). The term domain described in Section~\ref{sec:symbolic} is constructed analogously.
Right: A lifter ensures that both arguments to an operation are in the same domain.}\label{fig:simpledom}
\end{figure}

Even though cross-domain conversions are necessary in the program, it is a major task of the proposed transformation to minimise their number. A natural approach that would minimise unwanted domain
transitions is to propagate abstract domains along the data flow of the program. That is, if an abstract instruction \texttt{a\_inst} is already in the program and its result \(a\) is also used as an
operand elsewhere, we prefer to lift all the users of \(a\) into the abstract domain of \texttt{a\_inst} (cf. Figure~\ref{fig:simpledom}, right), instead of lowering \(a\) into a set of concrete
values. This simple technique, which we call \emph{value propagation}, forms the core of our entire approach (see also Figure~\ref{fig:syntactic}). It is worth noting that this is particularly simple
to do for programs in (partial) SSA form\footnote{Again, this is true of \llvm{} bitcode -- it is already in a partial SSA form. This simplifies our prototype implementation somewhat.}, although the
variables which are not part of SSA are still somewhat challenging. Those are covered by the \emph{freeze} and \emph{thaw} operations, which are discussed in more detail in Section~\ref{sec:freeze}.

Given the above, a logical starting point is to pick an initial set of instructions that we wish to lift into an abstract domain. Those could be explicit \emph{lift} instructions placed in the program
by hand, they could be picked by static analysis, or could be the result of abstraction refinement. The abstract program can be then obtained by applying value propagation to this initial set of
abstract instructions.

\subsection{Constraints and Relational Domains}\label{sec:constraints}

The last important aspect of abstraction is its effect on control flow of the program. It is often the case that control flow depends on specific values of variables via conditional branching. The
condition on the branch is typically a predicate on some value, or a relationship among multiple values that appear in the program. If the involved values are, in fact, abstract values, it is quite
possible that \emph{both} results of the predicate or comparison are admissible and that the conditional branch can therefore go both ways. The way we deal with this in the transformation is that the
program makes a \emph{nondeterministic choice} on the direction of the branch. How this nondeterministic choice is implemented is again deferred to the execution environment. In any case, the choice
of direction provides additional information -- constraints -- on the possible values of variables (cf. Figure~\ref{fig:cond}).

We encode those constraints into \emph{assume} instructions: given an abstract value and the constraint, \emph{assume} computes the constrained value. Additionally, depending on the abstract domain,
it may be desirable to constrain values other than those directly involved in the comparison. Alternatively, relational domains may be able to encode constraint information themselves: this is in
particular the case in the \emph{term domain} which realises symbolic computation. Therefore, for the purposes of the present paper, simply inserting a single \emph{assume} instruction on each
outgoing edge of the conditional is sufficient.

\subsection{Summary}\label{summary}

In the above, we have set up abstraction in such a way that it fits into a transformation-based approach. In particular, we have separated \emph{syntactic} and \emph{semantic} abstraction and shown
how the former induces the latter. The proposed syntactic abstraction captures how the program is changed, while semantic abstraction captures the dynamic (execution) aspects of abstract
interpretation.

\section{Symbolic Computation}\label{sec:symbolic}

Now that we have described how to perform program abstraction as a transformation, the remaining task is to re-cast symbolic computation as an abstract domain. Fortunately, this is not very hard: the
abstract values in the domain are \emph{terms}, while the abstract instructions simply construct corresponding terms from their operands. In other words, symbolic computation is realised by a
\emph{free algebra} (that is, the \emph{term algebra}). The \emph{input values} of the program correspond to nullary symbols -- in practice, a unique nullary symbol is created each time the program
obtains a value from its input. All the remaining values are built up as terms of bit-vector operations and constants. We will refer to the abstract domain thus formed as the \emph{term domain}.

It is not hard to see that a program transformed this way will simply perform part of its computation symbolically in the usual sense. Additionally, as the computation progresses, \emph{assume}
instructions impose a collection of \emph{constraints} on the nullary symbols of the abstract algebra (i.e.~the input values). Each constraint takes the form of a term with a relational symbol in the
root position. These constraints become part of the abstract state, effectively ensuring that the term domain is fully relational.\footnote{An abstract domain is called \emph{relational} when it is
  capable of preserving information about relationships among various abstract values that appear in the program.}

It is a requirement of abstract interpretation that it is possible to construct an abstract state from a set of concrete states. In the \emph{term domain} this can be achieved by assigning, to each
memory location that differs in some of the concrete states\footnote{In the present paper, we only deal with abstract (symbolic) \emph{values}. The structure of the program state, that is, the
  arrangement of the program memory, is taken to be always represented explicitly, i.e., it belongs squarely to the concrete domain.}, a fresh nullary symbol. We then impose constraints that ensure
that exactly the input set of concrete states is represented by the resulting abstract state. For instance, if the input set of concrete states differs by the value of a single variable \(a\), and
this variable takes values \(1\), \(2\), \(3\) and \(4\) in the 4 input states, a suitable constraint would be \(a \geq 1 ∧ a \leq 4\).

In some cases, it is impossible to construct the requisite constraints using only conjunction and relational operators. To ensure that the term domain forms a lattice (in particular that a least upper
bound always exists), it is necessary to allow the constraints to use logical disjunction.

While the above considerations regarding constraints are an important part of the theoretical underpinnings of the approach, it is almost always entirely impractical to shift back and forth between
concrete and abstract states. In practice, therefore, the constraints described in this section simply arise through the \emph{assume} mechanism described in Section~\ref{sec:constraints}. As such,
the constraints that appear in a given state form a \emph{path condition}. Finally, we note that the least upper bound of abstract states defined above corresponds to path conditions which arise from
\emph{path merging} in symbolic execution.

\section{Implementation}\label{sec:implementation}

We have implemented the proposed program transformation on top of \llvm{}, using its C++ API. Both the transformation and all additional code (model checker and solver integration) was done in C++. The
transformation itself is the largest component, totalling 3200 lines of code.

\subsection{Freeze and Thaw}\label{sec:freeze}

As mentioned in Section~\ref{sec:domains}, our implementation is based on the simple idea of maximum propagation of abstract values along the data flow of the program. While the SSA part of the
algorithm is essentially trivial, storing abstract values in program memory is slightly more challenging. The purpose of \emph{freeze} and \emph{thaw} is to overcome this issue.

While the dynamic type system that \emph{freeze} and \emph{thaw} provide to the transformed program and the ability to store additional data associated with a given memory address are largely
orthogonal at the conceptual level, they are closely related at the level of implementation. This is because in principle, a dynamic type system only requires that additional information is attached
to values manipulated by the program, and that this information is correctly propagated. Since apart from memory access, the program is statically typed, it is sufficient to perform dynamic type
checks (and dispatch) when a value is \emph{thawed}, while \emph{freeze} simply stores the incoming static type.

Implementation-wise, our target platform is a virtual machine with provisions for associating user-defined metadata to arbitrary memory addresses. This makes the implementation of \emph{freeze} and
\emph{thaw} simple and efficient. However, in case such a mechanism is not available, it is sufficient to implement an associative map, using addresses as keys, inside the program.

\subsection{Domains}\label{sec:impldomains}

In real-world programs, there are often variables which do not benefit from abstraction or from symbolic treatment, and are best represented explicitly. For this reason, the toplevel abstract domain
that we use is the disjoint union (i.e.~the type-level sum) of the concrete domain and the term domain. If we denote the concrete domain with \(\mathcal{C}\) and the symbolic (term) domain with
\(\mathcal{S}\), the type toplevel type is \(\mathcal{C} \sqcup \mathcal{S}\).

Since the \emph{freeze} and \emph{thaw} operations maintain dynamic type information in the executing program, it is possible to quickly compute operations for which both operands are concrete
(explicit). If both operands are symbolic, a symbolic operation is directly invoked, while in the remaining case -- one symbolic and one concrete argument -- the concrete argument is lifted into the
symbolic (term) domain. The procedure is called a \emph{lifter} and is automatically synthesized for each abstract operation that appears in the program. An example of a lifter is given in
Figure~\ref{fig:simpledom} (right).

It is also possible to use the domain \(\mathcal{C} \sqcup (\mathcal{C} × \mathcal{S})\), which corresponds to concolic execution (i.e.~it maintains both a concrete and a symbolic value at the same
time). This requires the additional provision that \emph{assume} instructions obtain concrete values that satisfy the symbolic constraints on their abstract counterparts (an SMT solver will typically
provide a model in case the assumptions were feasible, which can then be used to reconstruct the requisite concrete values).

\begin{figure}
\centering
\includegraphics{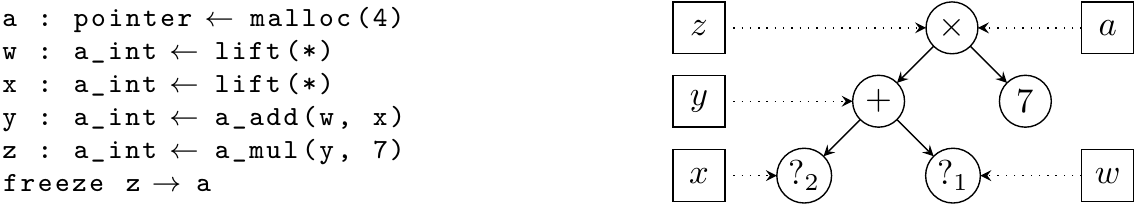}
\caption{Example of a formula tree as generated by the term domain. The boxes correspond to abstract variables, while the circles are the concrete representation of terms. Question marks denote
unconstrained nullary symbols.}\label{fig:tree}
\end{figure}

\begin{figure}
\centering
\includegraphics{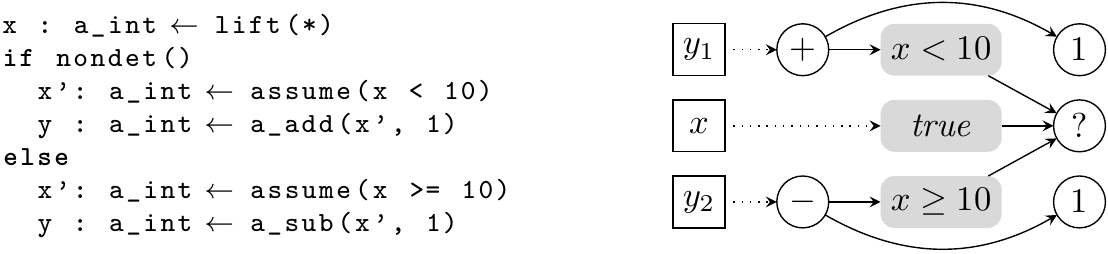}
\caption{The program on the left arises from instrumentation of conditional branching, in this case \texttt{if\ x\ \textless{}\ 10}. The formula tree on the right includes constraints arising from the
\emph{assume} instructions. Note that on any given path through the program, only one of the subtrees rooted in \(y₁\) or \(y₂\) can exist.}\label{fig:cond}
\end{figure}

\subsection{Execution \& Model Checking}\label{sec:symmc}

\begin{figure}
\centering
\includegraphics{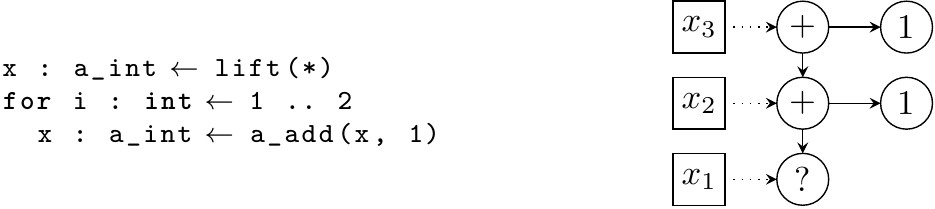}
\caption{An example of a formula tree arising from a \texttt{for} loop. Versions of the variable \texttt{x} which exist in different iterations of the loop are distinguished by an index in the
picture.}\label{fig:cycle}
\end{figure}

We represent the terms described in Section~\ref{sec:symbolic} by a simple tree data structure. The abstract instructions that correspond to operations on values construct a tree representation of the
requisite term by joining their operands to a new root node, where only the operation in the root node depends on the specific abstract instruction. The approach is illustrated in
Figure~\ref{fig:tree}, \ref{fig:cond}, \ref{fig:cycle}.

This arrangement makes it easy to extract the terms from program state and convert them to a form appropriate for further processing by the analysis tool. Recall that one of the motivating
applications of the proposed approach was symbolic model checking. In this case, the state space is explored by an explicit-state model checker and the extracted terms are converted into SMT queries.
To this end, the model checker must be slightly extended and coupled to an SMT solver, since:

\begin{enumerate}
\def\labelenumi{\arabic{enumi}.}
\tightlist
\item
  transitions of the program must be checked for \emph{feasibility},
\item
  the state equality check must compare terms semantically, not syntactically.
\end{enumerate}

Of course, the hitherto extracted terms must be left out of byte-wise comparison that is performed on the remaining (concrete) parts of program states. In our case, the required changes in the model
checker were quite minor, amounting to about 1200 lines of code.

\subsection{Interfaces}\label{sec:interfaces}

One of the goals of the proposed approach was to minimise interfaces between the abstracted program and the verification or execution environment (recall goal 2 set in Section~\ref{sec:goals}). In
total, there are four interactions at play:

\begin{enumerate}
\def\labelenumi{\arabic{enumi}.}
\tightlist
\item
  non-deterministic choice: under abstraction, conditionals in the program may be undetermined, and both branches may need to be explored; the abstraction uses a non-deterministic choice operator to
  capture this effect and defers an exploration strategy to the verifier
\item
  \emph{freeze} and \emph{thaw} must be provided as an interface for storing abstract values in program memory
\item
  enumeration of enabled (feasible) transitions must take the abstract values into account, if required by the domain(s) used in the program
\item
  state equality (if applicable in the verification approach) must be extended to take the employed abstract domains into account
\end{enumerate}

The latter two points depend on the chosen abstract domains. For the term domains, both interfaces reduce to extracting abstract values (terms) from program state and executing an SMT query.

\section{Evaluation}\label{evaluation}

First of all, we have checked the performance of the transformation itself. On C programs from the SV-COMP suite, the transformation time was negligible. On more complex C++ programs, it took at most
a few seconds, which is still fast compared to subsequent analysis.

As described in Section~\ref{sec:implementation}, we have built a simple tool which integrates the proposed transformation with an explicit-state model checker and an SMT solver. The experimental
evaluation was done using this prototype integration (denoted `\divine{}*' in summary tables).

\subsection{Code Complexity}\label{code-complexity}

One of our criteria for the approach presented in this paper was reduced code complexity. While counting lines of code is not a very sophisticated metric, it is a reasonably good proxy for complexity
and is easily obtained.\footnote{We have used the utility \texttt{sloccount} to get estimates of module size in terms of lines of code.} The results are summarised in Table~\ref{tbl:loc}.

\hypertarget{tbl:loc}{}
\begin{longtable}[]{@{}lrrrr@{}}
\caption{\label{tbl:loc}Summary of component sizes (thousands of lines of code) in a few symbolic verification and symbolic execution tools. Numbers in parentheses represent shared code (i.e.~code not
specific to the given approach to symbolic computation). }\tabularnewline
\toprule
component & \divine{}* & KLEE & Sym\divine{} & CBMC\tabularnewline
\midrule
\endfirsthead
\toprule
component & \divine{}* & KLEE & Sym\divine{} & CBMC\tabularnewline
\midrule
\endhead
transformation & 3.2 & 0 & 0 & (22)\tabularnewline
virtual machine & (10) & 15 & 6 & 7.5\tabularnewline
exploration & (1.5) & 1.2 & 1 & 2.3\tabularnewline
solver integration & 1.2 & 8 & 0 & 14\tabularnewline
SAT solver & (45) & (45) & (23) & (5.5)\tabularnewline
SMT solver & (80) & (80) & (400) & 16\tabularnewline
runtime support & 1 & 0 & 0 & 0\tabularnewline
\textbf{total} unique & 5.4 & 24.2 & 7 & 39.8\tabularnewline
\textbf{total} shared & 136.5 & 125 & 423 & 27.5\tabularnewline
\bottomrule
\end{longtable}

\subsection{Benchmarks}\label{sec:benchmarks}

For benchmarking, we have used a subset of the SV-COMP~\citep{beyer16:reliab.reprod} test cases, namely 7 categories, summarised in Table~\ref{tbl:svcomp}, along with statistics from our prototype
tool. We have only taken examples with finite state spaces since the simple approach outlined in Section~\ref{sec:symmc} cannot handle infinite recursion or infinite accumulation loops. In total, we
have selected 1160 SV-COMP inputs. In many cases (especially in the \texttt{array} category), the benchmarks are parametric: we have included both the original SV-COMP instance and smaller instances
to check that the approach works correctly, even if it takes a long time or exceeds the memory limit on the instances included in SV-COMP.

In all cases, the time limit, for each test case separately, was 10 minutes (wall time) and the memory limit was 10\,GiB. The test machines were equipped with 4 Intel Xeon 5130 cores clocked at 2\,GHz
and 16\,GiB of RAM.

In addition to the present approach, we have measured two additional tools: CBMC 5.8 and Sym\divine{}, both of which are symbolic model checkers targeting C code. The overall results of the comparison,
in terms of the number of cases solved, are presented in Table~\ref{tbl:solved}.

\subsection{Comparison 1: CBMC}\label{comparison-1-cbmc}

The results from CBMC 5.8 were obtained using the tool's default configuration. CBMC~\citep{kroening14:cbmc} is a mature bounded model checker for C programs with a good track record in SV-COMP and is
built around a symbolic interpreter for `goto programs', its own intermediate form, not entirely dissimilar to CIL or \llvm{} in its spirit. Besides KLEE, the CBMC toolkit is among the best established
members of the interpretation-based school of symbolic computation.

\hypertarget{tbl:svcomp}{}
\begin{longtable}[]{@{}lrrrrr@{}}
\caption{\label{tbl:svcomp}Summary of test cases from SV-COMP. The time limit was 10 minutes and memory limit was 10\,GiB. The oot/oom column is the number of test cases that were executed, but did
not finish within the given resource constraints. The `states' column gives the number of states stored, `search' gives the state space exploration time and `ce' gives the counterexample generation
time. }\tabularnewline
\toprule
tag & solved & oot/oom & states & search & ce\tabularnewline
\midrule
\endfirsthead
\toprule
tag & solved & oot/oom & states & search & ce\tabularnewline
\midrule
\endhead
array & 96 & 94 & 170.3 k & 52:00 & 54:15\tabularnewline
bitvector & 17 & 15 & 3166 & 3:12 & 2:33\tabularnewline
loops & 72 & 106 & 14.0 k & 53:52 & 11:40\tabularnewline
product-lines & 336 & 239 & 20.2 M & 4:36:44 & 43:11\tabularnewline
pthread & 9 & 36 & 609.4 k & 3:31 & 0:54\tabularnewline
recursion & 47 & 34 & 3955 & 16:16 & 7:41\tabularnewline
systemc & 14 & 45 & 25.0 k & 3:29 & 1:34\tabularnewline
\textbf{total} & 591 & 569 & & &\tabularnewline
\bottomrule
\end{longtable}

\hypertarget{tbl:solved}{}
\begin{longtable}[]{@{}lrrrr@{}}
\caption{\label{tbl:solved}The number of benchmarks correctly solved by each of the evaluated tools. The best result in each category is rendered in boldface. }\tabularnewline
\toprule
tag & total & \divine{}* & Sym\divine{} & CBMC\tabularnewline
\midrule
\endfirsthead
\toprule
tag & total & \divine{}* & Sym\divine{} & CBMC\tabularnewline
\midrule
\endhead
array & 190 & \textbf{96} & 68 & 93\tabularnewline
bitvector & 32 & \textbf{17} & 9 & 2\tabularnewline
loops & 178 & \textbf{72} & 67 & 9\tabularnewline
product-lines & 575 & 336 & \textbf{411} & 234\tabularnewline
pthread & 45 & \textbf{9} & 0 & 1\tabularnewline
recursion & 81 & \textbf{47} & 43 & 22\tabularnewline
systemc & 59 & 14 & \textbf{27} & 0\tabularnewline
\textbf{total} & 1160 & 591 & \textbf{625} & 361\tabularnewline
\bottomrule
\end{longtable}

Besides the total number of test cases solved (within the 10 minute limit), we were interested in comparing the time required to do so. Time requirements are summarised in Table~\ref{tbl:time}.

With regards to its state space exploration strategy, CBMC can be thought of as the middle ground between the approach taken by KLEE and that of our proposed tool. On one hand, KLEE, being a symbolic
executor, does not attempt to identify already-visited program states. CBMC is a bounded model checker, which means it stores a single formula representing the entire set of reachable states. Our
present approach, being based on an explicit-state model checker, stores sets of program states and compares them for equality using an SMT solver.

\hypertarget{tbl:time}{}
\begin{longtable}[]{@{}llllllll@{}}
\caption{\label{tbl:time}Speed comparison: the columns `models₁' and `models₂' show the number of models which the respective pair of tools finished in common. In most cases, CBMC is substantially
faster than the proposed approach, while Sym\divine{} is significantly slower. }\tabularnewline
\toprule
tag & models₁ & \divine{}* & CBMC & & models₂ & \divine{}* & Sym\divine{}\tabularnewline
\midrule
\endfirsthead
\toprule
tag & models₁ & \divine{}* & CBMC & & models₂ & \divine{}* & Sym\divine{}\tabularnewline
\midrule
\endhead
array & 73 & 34:16 & 13:58 & & 58 & 3:18 & 42:54\tabularnewline
bitvector & 2 & 0:37 & 0:01 & & 9 & 0:55 & 2:30\tabularnewline
loops & 4 & 0:03 & 0:02 & & 62 & 22:25 & 19:04\tabularnewline
product-lin. & 182 & 4:08:24 & 7:25 & & 183 & 0:30 & 28:33\tabularnewline
pthread & 0 & 0 & 0 & & 0 & 0 & 0\tabularnewline
recursion & 22 & 0:01 & 0:13 & & 43 & 4:02 & 13:58\tabularnewline
systemc & 0 & 0 & 0 & & 14 & 3:29 & 6:43\tabularnewline
\bottomrule
\end{longtable}

\subsection{Comparison 2: Sym\divine{}}\label{comparison-2-symdivine}

Sym\divine{}~\citep{mrazek16:symdiv} is a pre-existing, interpretation-based symbolic model checker which also works with \llvm{} bitcode. Similar to our approach, Sym\divine{} relies on a state equality
checker, in this case based on quantified bitvector formulae. In theory, this yields coarser state equivalence and consequently smaller state spaces, but we could not confirm this in our set of
benchmarks: the total number of states stored across the benchmarks that finished using both tools was 802 thousand for Sym\divine{} and 93 thousand with the approach described in this paper.
Additionally, QBV satisfiability queries are typically much more expensive than those used by our prototype tool, which can help explain the speed difference between the tools.

\section{Conclusion}\label{conclusion}

We have presented an alternate approach to symbolic execution (and abstract interpretation in general), based on compilation-based techniques, instead of relying on the more traditional
interpreter-based approach. We have shown that the proposed approach has important advantages and no serious drawbacks. Most importantly, our technique is modular to a degree not possible with
symbolic or abstract interpreters. This makes implementation of software analysis and verification tools based on symbolic execution almost trivial. An important side benefit is that the approach
allows for abstract domains other than the term domain, leading to a different class of verification algorithms with a comparatively small investment.

\bibliography{common}

\end{document}